\documentclass[aps,prl,twocolumn,showpacs,preprintnumbers,amsmath,amssymb]{revtex4}

\usepackage{undertilde}
\newcommand\be{\begin{eqnarray}}
\newcommand\ee{\end{eqnarray}}
\newcommand{\C}{{\mathbb C}}
\newcommand{\SOC}{{\rm SO}(3,\C)}
\newcommand\im{{\rm i}}

\begin{document}


\title{Metric Lagrangians with two propagating degrees of freedom}

\author{Kirill Krasnov}
  \affiliation{School of Mathematical Sciences, University of Nottingham, Nottingham, NG7 2RD, UK}

\date{October 21, 2009}

\begin{abstract} 
There exists a large class of generally covariant metric Lagrangians that contain only local terms 
and describe two propagating degrees of freedom. Trivial
examples can be be obtained by applying a local field redefinition to the Lagrangian
of general relativity, but we show that the class of two propagating degrees of freedom Lagrangians
is much larger. Thus, we exhibit a large family of non-local field redefinitions
that map the Einstein-Hilbert Lagrangian into ones containing only local terms. These 
redefinitions have origin in the topological shift symmetry of BF theory, to which GR is related 
in Pleba\'nski formulation, and can be computed order by order as expansions in powers of the 
Riemann curvature. At its lowest non-trivial order such a field redefinition produces 
the $(Riemann)^3$ invariant that arises as the two-loop quantum gravity counterterm. Possible 
implications for quantum gravity are discussed.
\end{abstract}


\maketitle

Loop divergences in quantum gravity require higher derivative counterterms
to be added to the Lagrangian \cite{'tHooft:1974bx}, \cite{Goroff:1985th}. Such higher 
derivative terms typically introduce new propagating degrees of freedom (DOF) that generally lead 
to instabilities, see \cite{Woodard:2009ns} for an emphasis of this point. The only known way to 
avoid these instabilities is to have a well-behaved underlying theory describing the new DOF, for 
example string theory. The higher-derivative metric Lagrangian is then an 
effective one obtained by integrating out some underlying non-gravitational DOF. In this letter we 
show that there exists a potentially attractive alternative: higher derivative counterterms 
can be added to the gravitational Lagrangian without adding new degrees of freedom. 

Field redefinitions play an important role in our construction, so we start by briefly
recalling some relevant facts. Quantum gravity, 
with its negative mass dimension coupling constant, is non-renormalizable in the sense 
that an infinite number of counterterms is required to absorb all arising divergences. 
However, while in a typical renormalizable theory transformations that absorb infinities are 
limited to field and coupling constant multiplicative renormalizations, the field redefinition 
freedom available in a theory with a dimensionful coupling is considerably larger. Thus, in the 
case of (pure, i.e with no matter couplings) quantum gravity one can perform field redefinitions of 
the schematic type $h\to h+ \sum_n G^n\partial^{2n} h+\ldots$,
where $h$ is the graviton field, $G$ is Newton's constant, and dots denote terms of higher order
in the metric perturbation. The power of $G$ here is as relevant for the case of 4 spacetime 
dimensions, but similar field redefinitions, with an appropriate modification of the power of $G$ 
are available in other dimensions as well. Such field redefinitions, being local, are known not
to change the S-matrix of the theory, see e. g. \cite{Marcus:1984ei}, section 2,
as well as \cite{'tHooft:1973pz}, section 10 for a discussion of this point. The availability of
these field redefinitions implies that many of the arising counterterms are unphysical
in the sense that they can be disposed off without any effect on the S-matrix.  An extreme
example of this situation arises when, in spite of divergences being present, they can  
all be removed by local field redefinitions without affecting the S-matrix. In this case one says 
that the theory is (on-shell) finite. An example of a finite but power-counting 
non-renormalizable theory is given by pure quantum gravity in 3 spacetime dimensions. 

For later purposes we note that classically a local metric field redefinition maps the
Einstein-Hilbert Lagrangian into a complicated metric Lagrangian containing an infinite number of
local terms. The new Lagrangian, however, still describes just two propagating DOF. 
This can be seen by following the Ostrogradski method of introducing new variables for 
higher time derivatives. One then observes that the arising Lagrangian
is highly degenerate and generates many constraints that remove all
but the DOF of the original system. It is simplest to see this mechanism at work by
considering a higher derivative field redefinition applied to a finite-dimensional 
dynamical system. 

A celebrated result of \cite{'tHooft:1974bx} is that one-loop divergences of pure quantum
gravity in 4 spacetime dimensions can be removed by a local field redefinition and so the
theory is one-loop finite. It was for some time hoped that
the finiteness may persist to all loops, but an explicit two-loop computation \cite{Goroff:1985th}
showed that the term $(Riemann)^3$ that is not removable by a local field 
redefinition is needed to absorb the divergences arising.

On the other hand, non-local redefinitions, i.e. involving negative powers of 
$\Box=\partial_\mu\partial^\mu$, generically do change the S-matrix. Still, an appropriate 
ghost action can be introduced to offset their effect, see \cite{'tHooft:1973pz}. However, such 
field redefinitions typically map a local action to a non-local one, and are thus uninteresting 
for the purpose of eliminating local counterterms. Indeed, the simplest example
is given by the free field Lagrangian $(1/2)(\partial_\mu\phi)^2$, which, after
a redefinition $\phi\to\phi+({\cal O}/\Box)\phi$, where ${\cal O}$ is some local operator,
goes into a non-local Lagrangian containing $1/\Box$.

We now show that in the case of gravity (in 4 spacetime dimensions) the class of
field transformations that map a local Lagrangian into again a local one is much larger
than that consisting of local field redefinitions. In other words, there exists an
(infinite-parameter) family of non-local field redefinitions that map the Einstein-Hilbert
Lagrangian into a generally covariant metric Lagrangian containing only local terms. 
The redefinition can be computed order by order in perturbation theory as follows. 
At lowest order, it is the local transformation
\be\label{h-loc}
h_{\mu\nu}\to h_{\mu\nu} +\alpha R_{\mu\nu} + \beta \eta_{\mu\nu} R
\ee
that produces $R^{\mu\nu}R_{\mu\nu}$ and $R^2$ invariants. Here $\eta_{\mu\nu}$
is the Minkowski metric, and $R_{\mu\nu}, R$ are the Ricci tensor and scalar for the
perturbation $h_{\mu\nu}$ respectively. At the next order our field redefinition 
produces the $(Riemann)^3$ invariant as well as other on-shell vanishing ones and is given by
\be\label{h-non-loc}
h_{\mu\nu}\to h_{\mu\nu} + \frac{\gamma}{\Box}  \partial^\alpha \partial^\beta 
R_{\mu\alpha}^{\quad\gamma\delta} R_{\nu\beta\gamma\delta},
\ee
plus a set of local terms. The reason why (\ref{h-non-loc}) produces
\be\label{two-loop}
\frac{\gamma}{4}\int R_{\mu\nu}^{\quad\rho\sigma}R_{\rho\sigma}^{\quad\alpha\beta}
R_{\alpha\beta}^{\quad\mu\nu}
\ee
is that this quantity can be written as:
\be\label{riemm-3}
\int {\cal E}^{\mu\nu}
\frac{\gamma}{\Box} \partial^\alpha \partial^\beta \left(
R_{\mu\alpha}^{\quad\gamma\delta} R_{\nu\beta\gamma\delta}
-\frac{1}{2} \eta_{\mu\nu}R_{\alpha}^{\quad\rho\gamma\delta} R_{\beta\rho\gamma\delta} \right),
\ee
where ${\cal E}_{\mu\nu}= R_{\mu\nu} - \frac{1}{2}\eta_{\mu\nu} R$.
This is checked using the easily verifiable identity
\be\label{ddR-ident}
\partial_{[\alpha}\partial^{[\beta} R_{\mu]}^{\,\,\nu]} = \frac{1}{4} \Box R_{\alpha\mu}^{\quad\beta\nu}
\ee
that holds to first order in the perturbation field. Note that the reason why the last term in
brackets in (\ref{riemm-3}) was not included in (\ref{h-non-loc}) is that it is proportional to 
$\eta_{\alpha\beta}$, and thus gives rise to a local term. 

The structure of the 
field redefinition at higher orders is similar to (\ref{h-non-loc}) in that the non-local operator 
$\partial^\alpha\partial^\beta/\Box$ is applied to a rank 4 tensor constructed from
an appropriate power of the Riemann curvature tensor (and its covariant derivatives), 
plus a set of local terms. Importantly, at higher orders there are also terms containing
higher negative powers of $\Box$. These are needed to eliminate terms arising as
powers of lower order non-localities.  The above prescription can be carried out order by
order, but this becomes technically difficult at higher orders. Below
we present an alternative description of the same field redefinition that guarantees
that it can be extended to any order and gives an algorithmic procedure for computing it. 

At every order the non-local field redefinition sketched introduces a set of parameters that,
after it is applied to the Einstein-Hilbert Lagrangian, translate into parameters of the arising
local metric Lagrangian. When truncated to any given order, the Lagrangian one obtains 
contains many new DOF stemming from its higher-derivative nature. However, the complete
Lagrangian with its infinite number of local terms describes just two propagating DOF. To see
this we must introduce a different and at first unrelated description of this class of Lagrangians.

An alternative description of the two propagating DOF metric Lagrangians is 
provided by an infinite-parameter family of deformations of general relativity first described in 
\cite{Bengtsson:1990qg}, building upon works 
\cite{Capovilla:1989ac,Capovilla:1991kx,Capovilla:1992ep,Bengtsson:1990qh}.
One starts with an observation \cite{Capovilla:1989ac}
that (complexified) Einstein's general relativity can be rewritten as a generally-covariant theory 
of an $\SOC$ connection. This suggests generalizations,
leading to an infinite-parameter family \cite{Bengtsson:1990qg} of theories describing two 
propagating degrees of freedom (DOF) and containing GR. These two propagating DOF
gravity theories can be rewritten in metric terms and  can be shown to be obtainable from 
GR precisely by the above non-local field redefinitions. 

These deformations of GR can be 
described compactly as follows. Let $A^i, i=1,2,3$ be an $\SOC$ connection and 
$F^i=dA^i+(1/2)\epsilon^{ijk}A^j\wedge A^k$ be its curvature two-form. The action of the theory is
just the most general generally-covariant action that can be constructed for $A^i$. 
Thus, consider the 4-form $F^i\wedge F^j$. Choosing an arbitrary volume 4-form $(vol)$ 
we can write $F^i\wedge F^j=(vol)\Omega^{ij}$, with $\Omega^{ij}$ being defined only 
modulo rescalings $(vol)\to\alpha(vol), \Omega^{ij}\to(1/\alpha)\Omega^{ij}$. 
Introduce a scalar-valued function $f(X)$ of $3\times 3$ symmetric matrices $X^{ij}$ that is
$\SOC$-invariant $f(OXO^T)=f(X), O\in\SOC$, holomorphic, and homogeneous of
degree one $f(\alpha X)=\alpha f(X)$. This function can be applied to the
4-form $F^i\wedge F^j$ with the result being a 4-form $f(F^i\wedge F^j)=(vol)f(\Omega)$,
independent of which volume 4-form is used. Thus, we can write a generally-covariant 
and gauge-invariant action as follows:
\be\label{action-A}
S[A]=\int f(F^i\wedge F^j).
\ee
It can then be shown that for a generic $f(\cdot)$ this gives a theory that describes
2 (complex) propagating DOF. This can be seen by noting that the phase space of this
theory is parametrized by pairs (spatial projection of the connection, conjugate
momentum). The theory is diffeomorphism and gauge-invariant which means that
there are $4+3$ first-class constraints acting on the phase space. With the configuration
space being $3\times 3$ dimensional, this leaves 2 physical DOF. 
It can also be shown, see \cite{Capovilla:1989ac}, that general relativity belongs to 
the class (\ref{action-A}) with the function $f(\cdot)$ being the $\delta$-function
projecting onto:
\be
{\rm Tr}\Omega^2 = \frac{1}{2}({\rm Tr}\Omega)^2.
\ee
Note that the clause about $f(\cdot)$ being generic is important, for the Lagrangian
${\rm Tr}F\wedge F$, which is also in the class (\ref{action-A}), is a total divergence
and corresponds to a theory without propagating DOF. The
description of the theory given here is new, but can be shown to be 
equivalent to one given in \cite{Bengtsson:1990qg}. 

As was realized in \cite{Krasnov:2006du,Krasnov:2008fm}, the theory 
(\ref{action-A}) can be put into a form that makes the spacetime metric it describes more explicit. 
In the retrospect, this is done via a standard "duality" trick of introducing a set of new fields 
that are later taken to be fundamental, with the fields of the original formulation to be 
integrated out. The new fields in our case are two-form fields $B^i$ that are valued in the 
Lie-algebra of $\SOC$. The new action is given by:
\be\label{action-AB}
S[B,A]=\int B^i\wedge F^i - \frac{1}{2}V(B^i\wedge B^j).
\ee
Here $V(\cdot)$ is again a holomorphic, $\SOC$-invariant, homogeneous
function of order one so that it can be applied to the 4-form $B^i\wedge B^j$.
Integrating the two-form field $B^i$ out by solving its (algebraic) field equations one
gets back (\ref{action-A}) with $f(\cdot)$ being an appropriate Legendre
transform of $V(\cdot)$. One can now take the two-form field $B^i$ to be 
fundamental, and eliminate $A^i$ completely by solving its field equations that are algebraic. 
This converts (\ref{action-AB}) into a second-order theory for the two-form field $B^i$.

The spacetime metric described by the theory becomes almost manifest in the two-form field 
formulation (\ref{action-AB}). Thus, it can be shown that
the theory is about the spacetime (conformal) metric
with respect to which the set of two-forms $F^i$ (or, equivalently, $B^i$) is self-dual.
It is not hard to show that there is a unique such (conformal) metric, see e.g. 
\cite{Urbantke:1984eb}. Explicitly, this metric is given by:
\be\label{Urb}
\sqrt{-g} g_{\mu\nu} \sim \epsilon^{ijk} B^i_{\mu\alpha} B^j_{\nu\beta} B^k_{\rho\sigma} 
\tilde{\epsilon}^{\alpha\beta\rho\sigma}.
\ee
Introducing the conformal metric (\ref{Urb}), the action (\ref{action-AB}) can be explicitly 
rewritten in a second-order form as that of the metric plus a set of auxiliary non-propagating
fields. This is done by introducing a set of special self-dual "metric" two-forms 
$\Sigma^i$ that satisfy:
\be\label{metr}
\Sigma^i\wedge \Sigma^j\sim \delta^{ij}.
\ee
These forms are easily constructed by introducing a tetrad $\theta^I, I=(0,i)$ for the metric,
and taking the self-dual part of the two-form $\theta^I\wedge \theta^J$ given by:
\be
\Sigma^i=\im \theta^0\wedge \theta^i - \frac{1}{2}\epsilon^{ijk}\theta^j\wedge \theta^k.
\ee
It can be shown that the knowledge of two-forms that are self-dual and satisfy
(\ref{metr}) is equivalent to the knowledge of the metric. A general self-dual 
two-form $B^i$ can then be written as:
\be
B^i = b^i_j \Sigma^j,
\ee
where $b^i_j$ are arbitrary scalars. The theory (\ref{action-AB}) with the connection
$A^i$ eliminated via its field equations then becomes a second-order theory of the
metric described by $\Sigma^i$ and the non-propagating scalars $b^i_j$. A simple
phase space analysis shows that the theory contains only two propagating DOF. The
scalars can then be integrated out to produce a purely metric theory. 
This leads to a Lagrangian given by an infinite expansion
in terms of local curvature invariants, which describes two propagating DOF by construction.
The above discussion was phrased in
terms of complex spacetime metrics, but appropriate reality conditions can be
imposed, and the story repeats itself for real Lorentzian signature metrics. 

Thus, we have seen that among all generally-covariant local (i.e. containing only local 
terms) metric Lagrangians there is an infinite-parameter subset that describes, as GR, only two
propagating DOF. To see the non-local field redefinitions that relate these Lagrangians 
to GR  we note that in the formulation (\ref{action-AB}) the first 
BF term possesses a large symmetry $B^i\to B^i+ D\eta^i$, where
$\eta^i$ is a Lie-algebra valued one-form, and $D$ is the covariant derivative
with respect to the connection $A^i$. A subgroup of this symmetry group is formed
by spacetime diffeomorphisms. The second, potential term of the action is only
invariant under this diffeomorphism subgroup, and this is the reason why 
the above "topological shift" transformation is not a symmetry of the whole action.
This is also the reason why (\ref{action-AB}), unlike BF-theory, has propagating
DOF. 

The topological shift transformation described can be used to map (\ref{action-AB}) to the 
Einstein-Hilbert action plus a simple potential term for a set of scalars that are
decoupled from the metric. At the linearized level this was noted in \cite{Freidel:2008ku}. 
To see this for the full theory, we use the observation of \cite{Plebanski:1977zz}
that the Einstein-Hilbert Lagrangian can be written in BF form in terms of the
two-forms $\Sigma^i$ constructed from the metric. We then note that the topological shift symmetry
can be used to transform any two-form field $B^i$ into a "metric"
one $\Sigma^i=B^i+D\eta^i$ satisfying (\ref{metr}). A detailed
demonstration of this fact is beyond the scope of this letter, but it is not hard
to see that the number of parameters available in the one-form field $\eta^i$,
modulo diffeomorphisms and modulo the "gauge" $\eta^i\to\eta^i+D\phi^i$,
where $\phi^i$ is a Lie-algebra valued zero-form, matches precisely the number of 
"metricity" equations (\ref{metr}) to be satisfied. It can also be shown,
at least perturbatively around the Minkowski background, that the two-form $\Sigma^i$
arising this way is unique. This discussion shows that
the first BF-term of the action (\ref{action-AB}) can be written as the Einstein-Hilbert
one for the metric obtained from $B^i$ by the topological shift symmetry.
The second term in (\ref{action-AB}) then becomes a potential for
the non-propagating scalars contained in $B^i$. By their field equations these scalars are set
to a value corresponding to a minimum of the potential, and decouple, which
leaves one with the Einstein-Hilbert action (with a cosmological constant
whose value is given by the minimum of $V(\cdot)$) for the metric described by
$\Sigma^i$. This shows that there exists a field redefinition that maps (\ref{action-AB})
into the Einstein-Hilbert action. The field redefinition in question involves solving
a differential equation for the shift one-form parameter $\eta^i$, and is thus non-local.
It  can be computed order by order perturbatively expanding the metric(s) around the 
Minkowski background. Details will appear elsewhere. The end result is given by the 
transformation that was described in the beginning of this letter, with parameters
of the transformation related to those of the potential $V(\cdot)$.

To summarize, we have seen that the set of generally-covariant local metric Lagrangians
describing two propagating DOF is larger than the one obtainable from GR by local 
field redefinitions, and admits a very compact description (\ref{action-A}). It is obtainable 
from GR by special non-local field redefinitions that stem from the topological shift 
symmetry of the BF-part of the action (\ref{action-AB}).

Let us conclude by discussing what the existence of an (infinitely) large class of two propagating 
DOF metric Lagrangians may imply for the problem of quantum gravity. The fact that the 
$(Riemann)^3$ counterterm needed at two loops is contained in our two propagating DOF
Lagrangians suggests that it may be possible to device a renormalization scheme for gravity so 
that the counterterm-corrected Lagrangian remains within two DOF class at every order of 
perturbative expansion. For this to be possible the class of theories (\ref{action-A}) must be closed 
under renormalization, which appears plausible, since the Lagrangian in (\ref{action-A}) is just
the most general one compatible with gauge and diffeomorphism invariance. Such a 
renormalization scheme, if possible, would give a quantum theory of gravity with two propagating 
DOF, which would be in striking contrast with other quantum gravity scenarios (e.g. string theory)  
that typically introduce new DOF. 

If this scenario was possible, one would face a question about 
implications of the non-local topological shift symmetry described. 
While generically non-local field redefinitions do change the S-matrix, our redefinitions are
certainly of a very special nature since a local action is mapped again into a local one. Therefore, 
the general conclusion has to be carefully re-examined. Preliminary considerations suggest that 
our non-local transformations might not affect the S-matrix. If this was so, then all quantum 
divergences were disposable without affecting the S-matrix, and the quantum theory would be
finite. It is of considerable interest to see if this vision can be realized.

The story described is that for pure, i.e. not coupled to any matter
sources, gravity. Indeed, coupling to usual type matter essentially removes the
field redefinition freedom. However, similarly to how the one-loop finiteness  result 
\cite{'tHooft:1974bx} extends to special matter couplings provided by 
supergravity theories, our story may also be applicable to gravity coupled to at least certain
types of matter. This will be described elsewhere. 

This work was supported by an EPSRC Advanced Fellowship. The author is grateful to
Alexey Boyarsky for important suggestions on the draft, and to Yuri Shtanov for
many discussions on the subject.


\begin{thebibliography}{0}

\bibitem{'tHooft:1974bx}
  G.~'t Hooft and M.~J.~G.~Veltman,
  Annales Poincare Phys.\ Theor.\  A {\bf 20}, 69 (1974).
  
\bibitem{Goroff:1985th}
  M.~H.~Goroff and A.~Sagnotti,
  Nucl.\ Phys.\  B {\bf 266}, 709 (1986).
  
\bibitem{Woodard:2009ns}
  R.~P.~Woodard,
  arXiv:0907.4238 [gr-qc].


\bibitem{Marcus:1984ei}
  N.~Marcus and A.~Sagnotti,
  Nucl.\ Phys.\  B {\bf 256}, 77 (1985).
 
\bibitem{'tHooft:1973pz}
  G.~'t Hooft and M.~J.~G.~Veltman,
  NATO Adv.\ Study Inst.\ Ser.\ B Phys.\  {\bf 4}, 177 (1974).
  
\bibitem{Bengtsson:1990qg}
  I.~Bengtsson,
  Phys.\ Lett.\  B {\bf 254}, 55 (1991).

\bibitem{Capovilla:1989ac}
  R.~Capovilla, T.~Jacobson and J.~Dell,
  Phys.\ Rev.\ Lett.\  {\bf 63}, 2325 (1989).
  
\bibitem{Capovilla:1991kx}
  R.~Capovilla, T.~Jacobson and J.~Dell,
  Class.\ Quant.\ Grav.\  {\bf 8} (1991) 59.

\bibitem{Capovilla:1992ep}
  R.~Capovilla,
  Nucl.\ Phys.\  B {\bf 373}, 233 (1992).

\bibitem{Bengtsson:1990qh}
  I.~Bengtsson and P.~Peldan,
  Int.\ J.\ Mod.\ Phys.\  A {\bf 7}, 1287 (1992).
  
\bibitem{Krasnov:2006du}
  K.~Krasnov,
  arXiv:hep-th/0611182.

\bibitem{Krasnov:2008fm}
  K.~Krasnov,
  Class.\ Quant.\ Grav.\  {\bf 26}, 055002 (2009)
  [arXiv:0811.3147 [gr-qc]].

\bibitem{Urbantke:1984eb}
  H.~Urbantke,
  J.\ Math.\ Phys.\ {\bf 25}, 2321 (1984).

\bibitem{Freidel:2008ku}
  L.~Freidel,
  arXiv:0812.3200 [gr-qc].
  
\bibitem{Plebanski:1977zz}
  J.~F.~Plebanski,
  J.\ Math.\ Phys.\  {\bf 18}, 2511 (1977).


\end{thebibliography}
\end{document}